\documentclass[aps,10pt,pre,twocolumn,english,groupedaddress, superscriptaddress]{revtex4-2}
\usepackage[T1]{fontenc}
\usepackage[utf8]{inputenc}
\usepackage{gensymb}
\usepackage{textcomp}
\usepackage{amsmath,amsthm,mathtools}
\usepackage{amssymb}
\usepackage{graphicx}
\usepackage{babel}
\usepackage[normalem]{ulem}
\usepackage{url}
\usepackage{appendix}
\usepackage{algorithm}
\usepackage{algorithmic}
\usepackage{cancel}
\usepackage{svg}
\usepackage{dsfont}
\usepackage{float}
\usepackage{float}

\usepackage{color}
\definecolor{blue}{rgb}{0,0,1}
\definecolor{dgreen}{rgb}{0,0.5,0}

\newcommand{\eder}{\textcolor{black}}

\usepackage{pbox} 
\definecolor{dblue}{rgb}{0,0,1}
\newcommand{\revise}[1]{\textcolor{black}{#1}}

\begin{document}

\author{Eder Batista Tchawou Tchuisseu}
\affiliation{Institute of Thermomechanics, Academy of Science of the Czech Republic, 18200 Prague 8, Czech Republic}
\email{ederbtt@it.cas.cz}
\author{Eric-Donald Dongmo}
\affiliation{Department of Mechanical Engineering, College of Technology, University of Buea, Po. Box 63, Buea, Cameroon}
\affiliation{Laboratory of Modeling and Simulation in Bio-Engineering and Prototypes, University of Yaounde 1}
\author{Pavel Proch\'azka}
\affiliation{Institute of Thermomechanics, Academy of Science of the Czech Republic, 18200 Prague 8, Czech Republic}
\email{proch@it.cas.cz}
\author{Paul Woafo}
\affiliation{Laboratory of Modeling and Simulation in Bio-Engineering and Prototypes, University of Yaounde 1}

\author{Pere Colet}
\affiliation{Instituto de Física Interdisciplinar y Sistemas Complejos, IFISC (CSIC-UIB), Campus Universitat Illes Balears, E-07122 Palma de Mallorca, Spain}
\author{Benjamin Sch\"afer}
\email{benjamin.schaefer@nmbu.no}
\affiliation{School of Mathematical Sciences, Queen Mary University of London, London E1 4NS, United Kingdom}
\affiliation{Faculty of Science and Technology, Norwegian University of Life Sciences, 1432 Ås, Norway}

\title{Secondary frequency control stabilizing the voltage}
\title{Secondary frequency control stabilizing voltage dynamics}

\begin{abstract}
The ongoing energy transition challenges the stability of the electrical power system. Stable operation of the electrical power grid requires both the voltage (amplitude) and the frequency to stay within operational bounds. While much research has focused on frequency dynamics and stability, the voltage dynamics has been neglected. 
Here, we study frequency and voltage stability in the case of \revise{simple networks}
via linear stability and bulk analysis. In particular, our linear stability analysis of the network shows that the frequency secondary control guarantees the stability of a particular electric network. 
Even more interesting, while we only consider secondary frequency control, we observe a stabilizing effect on the voltage dynamics, especially in our numerical bulk analysis.
\end{abstract}

\maketitle

\section{Introduction}
The need for good and stable electricity is a current and urgent quest in our society \cite{Machowski2011,mancarella2021introduction}. Electricity is generated by the conversion of a primary source of energy such are mechanical, chemical, nuclear or thermal to electrical energy. To power small devices, energy harvesting systems are commonly used \cite{tchawou2014harvesting, tekam2014analysis}, while for powering cities or countries synchronous generators and renewable energy sources are needed \cite{tchuisseu2018curing, Witthaut2012, Tchawou1}. The infrastructure connecting such generators and consumers of electricity is called the electric power grid.
The traditional power system stability remains an important task to be achieved by electric utilities in order to ensure a good electric quality and supply security for the consumers \cite{tchuisseu2018curing, Witthaut2012, Witthaut2013, Schaefer2015, dongmo2015effects, Dongmo2017, Filatrella2008}. Power imbalances are one of the principal causes of grid instabilities, which can lead the network to blackouts. In fact, any power imbalance induces a variation of the frequency and the voltage (amplitude) of the electric grid.  

In the literature, much work has been devoted to propose controllers that stabilize the grid when facing any power imbalance. Many such controllers mainly focused on either controlling the frequency \cite{tchuisseu2018curing, Dongmo2017, Tchawou1, gorjao2020data} of the grid or the voltage \cite{sun2019review} of the power grid. For the frequency control, the governor of the power plant is often used through the load frequency control and the automatic generation control, mostly known as the primary, secondary and tertiary frequency controls \cite{Tchawou1,Weitenberg2017,tyloo2020primary,bottcher2020time}. The voltage on the other hand is controlled through the automatic voltage regulator, which ensures that the voltage is kept within an admissible range. Therefore, many studies only focused on either the control of the frequency or the control of the voltage and rarely both.

There are three leading models \cite{nishikawa2015comparative} to mathematically describe the  power grid, which are: the effective network; the structure-preserving model and the synchronous motor model. In this paper, we are using the synchronous motor model, where each node in the network ("the motor") can be considered as an aggregate of generators or consumers (e.g. a small region or large city). The voltage sources of these machines are usually considered constant, such that the dynamic of the power system is reduced to its frequency, hence phase dynamics \cite{nishikawa2015comparative, tchuisseu2018curing, Schaefer2016, Schaefer2017}. 

\revise{Within this article, we investigate the stability of the high-voltage transmission system while including the voltage dynamics. This model has been studied in the electric network research community e.g. by Katrin Schmietendorf et al. in \cite{schmietendorf2014}, Sabine Auer et al. \cite{auer2016} and Florian Döfler et al. \cite{Doerfler2014}, all stressing the need to include voltage dynamics even on transmission level.} However, these works do not consider any type of control applied to the variables of the network (phase, frequencies and voltages). But this control is essential for the global understanding of the power system dynamics and stability. For example, it has been shown that applying secondary frequency control to the well-known second order electric network model, can deeply modify the dynamics of a network \cite{Machowski2011,Weitenberg2017,tchuisseu2018curing}. 
Thus, we aim to couple the frequency secondary control to the electric network considering the voltage dynamics. The results obtained including the voltage dynamics will be compared with the ones obtained with the classical uncontrolled model.

The rest of this paper is organized as follows: Section 2 provides a mathematical model of the power grid considered in this work as networks of synchronous machines controlled each by frequency controllers.  Based on linear stability analysis and bulk dynamics, we also quantify the stability of the power grid. Numerical analyses are presented in Section 3 for small (N=2) and larger (N>2) networks. Finally, the paper is concluded in Section 4.

\section{Mathematical model and stability analysis}
\subsection{Mathematical model}

The electric network is modeled as coupled synchronous machines combined with their respective frequency controllers (secondary) and by considering the voltage dynamics, as follows:
\begin{equation}  
\left\{
    \begin{array}{ll}
      \dot{\mathbf{\theta}}_i=\omega_i\\
      \dot{\omega}_i=-\alpha_{i}\dot{\mathbf{\theta}}_i+P_{i}^{*}-\sum\limits_{j=1}^N E_{q,i}^{'}B_{ij}E^{'}_{q,j}\sin(\theta_{i}-\theta_{j})+ u_{i}\\
      T^{'}_{{d,i}_{0}}\dot{\overline{E}}_{q,i}^{'}=E_{f,i}-E_{q,i}^{'}+ (X_{d,i}-X_{d,i}^{'})\sum\limits_{j=1}^N B_{ij}E^{'}_{q,j}\cos(\theta_{i}-\theta_{j})\\
      \tau_{g_i}\dot{u}_i= -u_{i} -\gamma_{i}\theta_{i} - \beta_{i}\omega_i,
    \end{array}
\right.
 \label{full_3r_dmodel} 
\end{equation}
where $i\in {1,...,N}$ denotes the node index in the network, $\theta$ the voltage phase angle, $\omega$ the frequency, $E$ the voltage amplitude and $u$ the control. $\alpha$ is a damping constant, $P^*$ the power consumed/generated at a node, $B$ the susceptibility matrix, while $T$ and \eder{$\tau_{g}$} are time constants, \eder{$E_{f}$ is the rotor’s field voltage},  and $X_d$ and $X_d^{'}$ are voltage dynamic parameters \cite{schmietendorf2014}. The nodes (islanded power grids or synchronous machines), which compose the network, are assumed to be all to all connected.
For simplicity we assume that the frequency controller acts instantaneously \eder{($\tau_{g}$=0)}, such that it is described as the proportional derivative control given by: 

$u_{i} =-\gamma_{i}\theta_{i} - \beta_{i}\omega_i$. The derivative term $\beta_{i}\dot{\mathbf{\theta}}_{i}$ can be absorbed into the damping term $\alpha_{i}\dot{\mathbf{\theta}}_{i}$ of the swing equation.
 One can then also rewrite the set of equation~\ref{full_3r_dmodel} by Eq.~\ref{full_3r_dmodel2} substituting $(X_{d,i}-X_{d,i}^{'})$ by $X_{i}$, $T^{'}_{{d,i}_{0}}$ and $E_{q,i}^{'}$ by respectively 
 $T_{{d,i}}$ and $E_{i}$ without loss of generality
\begin{equation}  
\left\{
    \begin{array}{ll}
      \dot{\mathbf{\theta}}_i=\omega_i\\
      \dot{\omega}_i=-\alpha_{i}\dot{\mathbf{\theta}}_i-\gamma_{i}\theta_{i}+P_{i}^{*}-\sum\limits_{j=1}^N E_{i}B_{ij}E_{j}\sin(\theta_{i}-\theta_{j})\\
      T_{{d,i}}\dot{\overline{E}}_{i}=E_{f,i}-E_{i}+ X_{i}\sum\limits_{j=1}^N B_{ij}E_{j}\cos(\theta_{i}-\theta_{j}).\\
   \end{array}
\right.
 \label{full_3r_dmodel2} 
\end{equation}

\subsection{Linear stability analysis}
The stable operation of the electric network requires that the frequency, voltage and difference of phases between connected nodes is constant, meaning that the network is operating in a synchronous regime. Such system is said to be linearly stable if, subjected to a small perturbation, it regains its stable operation, in the case of a power grid its synchronous state. Konstantin Sharafutdinov et al. have extensively studied the stability of such electric network model considering the voltage dynamics as described in Eq.~\eqref{full_3r_dmodel2}, but without secondary control. Thus, based on their work and mainly on the necessary and sufficient conditions of the uncontrolled system to be linearly stable, the effects of the secondary control on the stability of the network will be investigated.\\

Thus, to analyze the stability of the system with respect to small perturbation, we linearize Eq.~\ref{full_3r_dmodel2} around a steady state $(\theta_{i}^{*},\omega_{i}^{*},E_{i}^{*})$. We denote small perturbation around the steady state as $\theta_{i}=\theta_{i}^{*}+\delta\theta_{i}$,
$\omega_{i}=\omega_{i}^{*}+\delta\omega_{i}$, $E_{i}=E_{i}^{*}+\delta E_{i}$. The linearization of the equation~\ref{full_3r_dmodel2} leads to the Eq.~\ref{voltlinear}, where $\mathbf{X_{1}}$, $\mathbf{X_{2}}$ and 
$\mathbf{X_{3}}$ are n-dimensional vectors of $\delta\theta_{i}$, $\delta\omega_{i}$ and $\delta E_{i}$.
\begin{equation}  
\left\{ \begin{array}{ll}
      \dot{\mathbf{X}}_\mathbf{_1}=\mathbf{X_2},\\
      \dot{\mathbf{X}}_\mathbf{_2}=-(\mathbf{P}+\mathbf{\Gamma})\mathbf{X_1}-\mathbf{A}\mathbf{X_2} - \mathbf{\Lambda}\mathbf{X_3} \\
      \dot{\mathbf{X}}_\mathbf{_3}=\mathbf{T}^{-1}\mathbf{\chi}\mathbf{\Lambda}\mathbf{X_1} + \mathbf{T}^{-1}(\mathbf{\chi}\mathbf{C}-\mathds{1})\mathbf{X_3},
     \end{array}
\right.
 \label{voltlinear} 
\end{equation}
where $\mathbf{T}^{-1}$, $\mathbf{\Gamma}$, $\mathbf{A}$ and $\mathbf{\chi}$ are diagonal matrices with elements $T_{ii}$= $\frac{1}{T_{i}}$, $\mathbf{\Gamma_{ii}}$= $\gamma_i$,$A_{ii}$= $\alpha_i$ and $\chi_{ii}$=$X_{i}$ 
respectively, representing the relaxation time of the transient voltage dynamics matrix,the control, the damping and the transient reactance matrix of the synchronous machine. Matrices $\mathbf{P}$, $\mathbf{\Lambda}$, $\mathbf{C}$, $\mathbf{\Gamma}$ $\in$ $\mathds{R}^{n \times n}$ whose the elements 
are respectively defined as follows:
\begin{equation}
P_{ij}=\left\{ \begin{array}{ll} 
		-{E_{i}}^{*}B_{ij}{E_{j}}^{*}\cos({\theta_{i}}^{*}-{\theta_{j}}^{*}), 
		& i \ne j,\\
        \sum\limits_{l \ne i}^N {E_{i}}^{*}B_{il}{E_{l}}^{*}\cos({\theta_{l}}^{*}-{\theta_{i}}^{*}),
        & i=j,
          \end{array}
\right.
\end{equation}
\begin{equation}
{\Lambda}_{ij}=\left\{ \begin{array}{ll} 
		{E_{j}}^{*}B_{ij}\sin({\theta_{i}}^{*}-{\theta_{j}}^{*}), 
		& i \ne j,\\
        -\sum\limits_{l= i}^N {E_{l}}^{*}B_{il}\sin({\theta_{l}}^{*}-{\theta_{i}}^{*}),
        & i=j,
          \end{array}
\right.
\end{equation}
\begin{equation}
{C}_{ij}=B_{ij}\cos({\theta_{i}}^{*}-{\theta_{j}}^{*}).
\end{equation}
This later set of equations can be rewritten into the following compact form

\begin{equation}
\frac{\text{d}}{\text{d}t}\left( {\begin{array}{ccc}
 \mathbf{X_{1}}\\
 \mathbf{X_{2}}\\
 \mathbf{X_{3}}
\end{array} } \right)
= \underbrace{\left[ {\begin{array}{ccc}
 \mathbf{0} & \mathds{1} & \mathbf{0} \\
 -(\mathbf{P}+\mathbf{\Gamma}) & -\mathbf{A} & - \mathbf{\Lambda}\\
 \mathbf{T}^{-1}\mathbf{\chi}\mathbf{\Lambda} & \mathbf{0} &\mathbf{T}^{-1}(\mathbf{\chi}\mathbf{C}-\mathds{1})
\end{array} } \right]}_{:=\mathbf{J}}\left[ {\begin{array}{cc}
 \mathbf{X_{1}}\\
 \mathbf{X_{2}}\\
 \mathbf{X_{3}}
\end{array} } \right].
\end{equation}

The evaluation of the Jacobian matrix $\mathbf{J}$ around an existing fixed point $(\theta_i^{*}, \omega_i^{*}, E_i^{*})$ determines the stability of the system around this fixed point. That is done by computing the eigenvalues $\mu_{i}$ associated to this fixed point. The system is then linearly stable if the maximal real part of the eigenvalues is negative. In this case, perturbed, the system will regain its fixed point. But, if at least one eigenvalue has a positive real part, then, the system is said to be linearly unstable. This previous analysis is possible once we can determine the steady state of the system and therefore compute the eigenvalues.\\
In most of the cases, the fixed points can be difficult to determine, hence complicating the determination of eigenvalues. In such cases, the determination of condition of existence of the fixed point is necessary and the stability conditions of the system can be derived through some mathematical formulations.\\
Sharafutdinov \textit{et al} in \cite{sharafutdinov2018rotor} have provided for the uncontrolled system two sufficient and necessary stability conditions grouped into two propositions (proposition I and II in \cite{sharafutdinov2018rotor}), which both have to hold to have a linearly stable fixed point. In fact, it is worth noting that the Jacobian matrix $\mathbf{J}$ has one eigenvector $(1,0,0)$ with eigenvalue $\mu_{1}= 0$, which corresponds to a global phase shift of the synchronous machine \cite{Manik2014,sharafutdinov2018rotor}.
This particular case, which does not have any physical implications for the stability, is excluded from the stability analysis. Thereby we reduce the space of possible states to  $S_{\bot}$, which is the space perpendicular to the solution {$\theta_i^{*}$+c(1,0,0)/c $\in \mathds{R}$} and defined by
\begin{equation} S_{\bot}=\left\{ (\mathbf{X_{1}},\mathbf{X_{2}},\mathbf{X_{3}})
\in \mathds{R}^{3n}/\mathbf{1}\mathbf{X_{1}}= 0\\
\right\}.
\end{equation}
The Proposition I in \cite{sharafutdinov2018rotor} states that a given steady state $(\theta_i^{*}, \omega_i^{*}, E_i^{*})$ is linearly stable if and only if 
\begin{itemize}
 \item given the space $S_{\bot}^{1}=\left\{ (\mathbf{X_{1}},\mathbf{X_{2}}) \in \mathds{R}^{2N}/\mathbf{1}\mathbf{X_{1}}= 0\\
\right\}$, the matrix $\mathbf{P+\Gamma}$ is positive definite on $S_{\bot}^{1}$, furthermore,
\item the matrix $\mathbf{C}-\mathbf{\chi}^{-1}+ \mathbf{\Lambda}\mathbf{P}^{+}\mathbf{\Lambda}^{T} $ is negative definite.
\end{itemize}
Where $T$ denotes the transpose of a matrix, which is identical to its inverse if it is an orthogonal matrix. In addition, the matrix $\mathbf{P}^{+}$, represents the Moore-Penrose pseudo-inverse of  $\mathbf{P}$. From this proposition I, the only difference appearing in comparison to the study made in Sharafutdinov \textit{et al} in \cite{sharafutdinov2018rotor} is the matrix $\mathbf{\Gamma}$ in the sum $\mathbf{P+\Gamma}$. Thus, from the proposition I, it is clear that the secondary control represented by the matrix $\mathbf{\Gamma}$ can improve the stability of the network with the necessary condition to be positive definite. In our case, where the nodes are assumed to have the same system parameters such are $\alpha_{i}=\alpha$, $\gamma_{i}=\gamma$, $T_{d,i}=T_{d}$ and $E_{f,i}=E_{f}$. The matrix  $\mathbf{\Gamma}$ is a diagonal matrix, which elements $\gamma$ are always positive. Thus, all the eigenvalues of the symmetric matrix  $\mathbf{\Gamma}$ are positive, hence, the matrix  $\mathbf{\Gamma}$  is positive definite (See Appendix~A).\\

\subsection{Bulk dynamics}
In order to analyze the stability of the network, we simplify the analysis by focusing on the ensemble dynamics, i.e. we consider the \emph{bulk} or average dynamics of the network. As in the previous stability analysis, we wish to evaluate the impact of the secondary control through its parameter $\gamma$ on the dynamics of the system. Thus, we  first consider a system with a constant voltage, which is a well known and studied case. Next, we include the voltage dynamics and analyze the effects of the secondary control on the voltage.

\subsubsection{Constant voltage}
Let' s consider Eq.~\eqref{full_3r_dmodel2} and assume the voltage to be constant. 
The resulting equation is the simple Kuramoto model with secondary control, discussed for example in \cite{tchuisseu2018curing}:

\begin{equation}  
\left\{
    \begin{array}{ll}
      \dot{\mathbf{\theta}}_i=\omega_i,\\
      \dot{\omega}_i=-\alpha_{i}\dot{\mathbf{\theta}}_i-\gamma_{i}\theta_{i}+P_{i}^{*}-\sum\limits_{j=1}^N E_{i}B_{ij}E_{j}\sin(\theta_{i}-\theta_{j}).\\
 \end{array}
\right.
 \label{full} 
\end{equation} 
Now, we take the average of the variables of Eq.~\eqref{full} over the number of nodes $n$ that form the network. We further assume that the node parameters $\alpha_i=\alpha$, $\gamma_i=\gamma$ are identical for all the nodes. Thus, we obtain the following differential equations:

\begin{equation}  
\left\{
    \begin{array}{ll}
      \dot{\overline{\theta}}=\overline{\omega},\\
      \dot{\overline{\omega}}=-\alpha\dot{\overline{\theta}}-\gamma\overline{\theta}+\frac{1}{N}\sum_{i=1}^N {P_{i}^{*}}.\\
 \end{array}
\right.
 \label{bulkeq1} 
\end{equation} 
Eq.~\eqref{bulkeq1} can easily be solved and the average value of the principal variable are obtained as follows: 

\begin{equation*}
    \overline{\theta}(t)=C_{1}\exp{r_1t}+ C_{2}\exp{r_2t}+\frac{\sum_{i=1}^N {P_{i}^{*}}}{N\gamma},
\end{equation*}
where $C_1$ and $C_2$ are constants which are determined by the initial conditions and $r_1$ and $r_2$ are expressed as follows:

\begin{equation}  
\left\{
    \begin{array}{ll}
      r_1=-\frac{\alpha}{2}+\frac{\sqrt{\alpha^2 - 4\gamma}}{2},\\
       r_2=-\frac{\alpha}{2}-\frac{\sqrt{\alpha^2 - 4\gamma}}{2}.\\
 \end{array}
\right.
\end{equation} 
Thus, it is clear that for time tending to infinity and for any value of the secondary control parameter $\gamma$ ($\gamma>0$), the average value of the angle of rotation of each node in the network is constant and the average value of the frequency is zero: 
\begin{equation*}
    \overline{\theta}(t)=\frac{\sum_{i=1}^N {P_{i}^{*}}}{N\gamma}, \quad \bar{\omega}=0. 
\end{equation*}
As we can see, the average of the angle $\overline{\theta}$ for the controlled system depends on the control parameter $\gamma$, the balance of the network as well as the size of the network. For a balanced network, i.e. with $\sum_{i=1}^N {P_{i}^{*}}=0$, the mean value of the angles is zero. For an imbalanced and fixed network, the mean angle is inversely proportional to the secondary control parameter, thus for a large control, the average of the angles tends to zero.\\
For an uncontrolled electric network ($\gamma=0$), the average of the angles is time varying and given by the following expression:
\begin{equation*}
    \overline{\theta}(t)=D_{1}+ D_{2}\exp{(-\alpha t)}+\frac{\sum_{i=1}^N {P_{i}^{*}}}{N\alpha}t,
\end{equation*}
where $D_1=-D_2=-{\sum_{i=1}^N{{P_{i}^{*}}}/{N\alpha^2}}$ are constants determined by the initial conditions. Thus, the mean value of the frequency of the network as a function of time $t$ is given by the Eq.~\eqref{dev}: 
\begin{equation}
   \overline{\omega}(t)= -\alpha{D_{2}}\exp{(-\alpha t)}+\frac{\sum_{i=1}^N {P_{i}^{*}}}{N\alpha},
   \label{dev}
\end{equation} 
this basically corresponds to the frequency deviation for an uncontrolled and imbalanced network.

This imbalance can be easily absorbed in large networks (large $N$) and those with large primary control (given by the $\alpha$ parameter).\\ 

We have so far shown that, when the voltage is constant, any imbalance in the network induces the deviation of the mean frequency, which is constant and non-zero for a network without control. This implies that the average value of the angle is linearly increasing with the time. For the same network, but in presence of secondary control in all the nodes, one notices that the average frequency in the network is always zero after a long period of time and the average angular in the network is a constant value proportional to the disturbance and inversely proportional to the size of the network and the secondary control parameter. We have therefore seen the effects of the secondary control in such an electric network in which the voltage dynamics is not considered. But how does the voltage dynamics change these results?

\subsubsection{Dynamical voltage}
In this part, we investigate the effects of the secondary control on the dynamics of the voltage in an electric network. 
Thus, we consider the full equation~\eqref{full_3r_dmodel2}, summing up all three equations and dividing them by the total number of nodes. Again, we assume that the parameters of the nodes are constant and identical for all the nodes, obtaining the following equations: 
\begin{equation}
\left\{
    \begin{array}{lll}
      \dot{\overline{\theta}}=\overline{\omega}\\
      \dot{\overline{\omega}}=-\alpha\dot{\overline{\theta}}-\gamma\overline{\theta}+\frac{1}{N}\sum_{i=1}^N {P_{i}^{*}}\\
    T_{d}\dot{\overline{E}}=E_f -\overline{E} + \frac{X}{N}\sum_{i=1}^N\sum\limits_{j=1}^N B_{ij}E_{j}\cos(\theta_{i}-\theta_{j})  
 \end{array}
\right.
\label{eq3} 
\end{equation}
As shown in Eq.~\eqref{eq3}, the equations describing the dynamics of the mean voltage (the third equation) contain the difference of phases of connected nodes as argument of a cosine function. The presence of this term makes the determination of the mean voltage fixed point very difficult. Nevertheless, one can find the range of variation of $\overline{E}$ considering the following relations:
\begin{equation*}
    \begin{cases}
    B_{ij}=B_{ji}=\left\{
    \begin{array}{ll}
        B_{0} & \mbox{if } i=j \\
        B_{1} & \mbox{else}
    \end{array}\right.\\
   -1 \leq \cos{(\theta_{i}-\theta_{j})}\leq 1.

    \end{cases}
\end{equation*}
Thus, the term with cosine in Eq.\eqref{eq3} is approximated by the following relation: 
\begin{equation}
\begin{cases}
-X(B_{0}+(N-1)B_{1})\overline{E}\leq \frac{X}{N}\sum_{i=1}^N\sum\limits_{j=1}^N B_{ij}E_{j}\cos(\theta_{i}-\theta_{j}), \\
\frac{X}{N}\sum_{i=1}^N\sum\limits_{j=1}^N B_{ij}E_{j}\cos(\theta_{i}-\theta_{j})\leq X(B_{0}+(n-1)B_{1})\overline{E}.
\end{cases}
\end{equation}
The range of variation of the mean value can then be determined, solving the following inequalities:
\begin{equation}
    \begin{cases}
T_{d}\dot{\overline{E}}+\overline{E}-E_f\leq X[\ (B_{0}+(N-1)B_{1})]\ \overline{E},\\
T_{d}\dot{\overline{E}}+\overline{E}-E_f \geq X[\ (B_{0}-(N-1)B_{1})]\ \overline{E}.
    \end{cases}
    \label{meanvolta}
\end{equation}
Thus, the average of the voltage is bounded as follows:
\begin{equation}
 \begin{cases}
  \overline{E(t)} \leq E_{0}\exp{(\frac{-(1-\sigma_1)t}{T_{d}})}+\frac{E_{f}}{T_{d}(1-\sigma_1)}(\ 1-\exp{(\frac{-(1-\sigma_1)t}{T_{d}})})\,\\
  \overline{E(t)} \geq E_{0}\exp{(\frac{-(1+\sigma_2)t}{T_{d}})}+\frac{E_{f}}{T_{d}(1+\sigma_2)}(\ 1-\exp{(\frac{-(1+\sigma_2)t}{T_{d}})})\ ,
 \end{cases}
 \label{ine}
\end{equation}
where $E_{0}$ is the initial value of the mean voltage: $\overline{E}(t=0)=E_{0}$, and $\sigma_2$ and $\sigma_1$ are given by:
\begin{equation*}
 \begin{cases}
    \sigma_1 =X[\ (B_{0}+(N-1)B_{1})],\ \\
    \sigma_2 =-X[\ (B_{0}+(N-1)B_{1})].\
     \end{cases}
\end{equation*}
Evaluating all previous inequalities, we observe that the dynamics of the mean voltage is strictly related to the size of the network. Thus, according to Eq.~\eqref{ine}, the mean voltage will be bounded between real values if and only if:
\begin{equation}
    \begin{cases}
   1-\sigma_1 =1-X[\ (B_{0}+(N-1)B_{1})]\geq 0, \\ 
   1-\sigma_2 =1+X[\ (B_{0}-(N-1)B_{1})] \geq 0.\\
   
    \end{cases}
\end{equation}
Thus the stability of the electric voltage is achieved if the network size and the network coupling are chosen in the range given by the following inequality: 
\begin{equation}
   1 -X\frac{1-B_{0}}{B_{1}}\leq N\leq 1+X\frac{1-B_{0}}{B_{1}}
\end{equation}
In the present study, the values of the system's parameters are set as: $X=1$, $B_{0}=-0.8$ and $B_{1}=1$\revise{, following literature values \cite{schmietendorf2014, auer2016}}. Thus, the mean voltage of the network of size $N$ will be bounded if and only if: \begin{equation}  \eder{N= 2}\end{equation}
Indeed, let's consider an electric network with the same parameters as the studied case and composed of $n$ nodes, where each node represents a small region or a city. The result above states that an extension of the current network composed of $N=2$ nodes by connecting it with additional nodes leads to the instability of the new extended network in terms of voltage. 

\revise{Thus, from this bulk analysis, it appears that the secondary frequency control has explicitly no effects on the stabilization of the mean voltage, but does have an effect when it comes to stabilizing the mean frequency of the network. This implies therefore the necessity of having another form of control for the mean voltage in larger electric networks.}

\section{Numerical analysis}
 In the previous section, we analysed the linear stability as well as the bulk dynamics analytically. In this part, we present a numerical analysis for N=2  and $N\geq2$ nodes with the objective to evaluate the effects of the frequency secondary control ($\gamma$) on the network, thereby complementing the analytical results. 
\subsection{2-node system}
\begin{figure}[ht]
\begin{centering}
\includegraphics[width=1.1\linewidth]{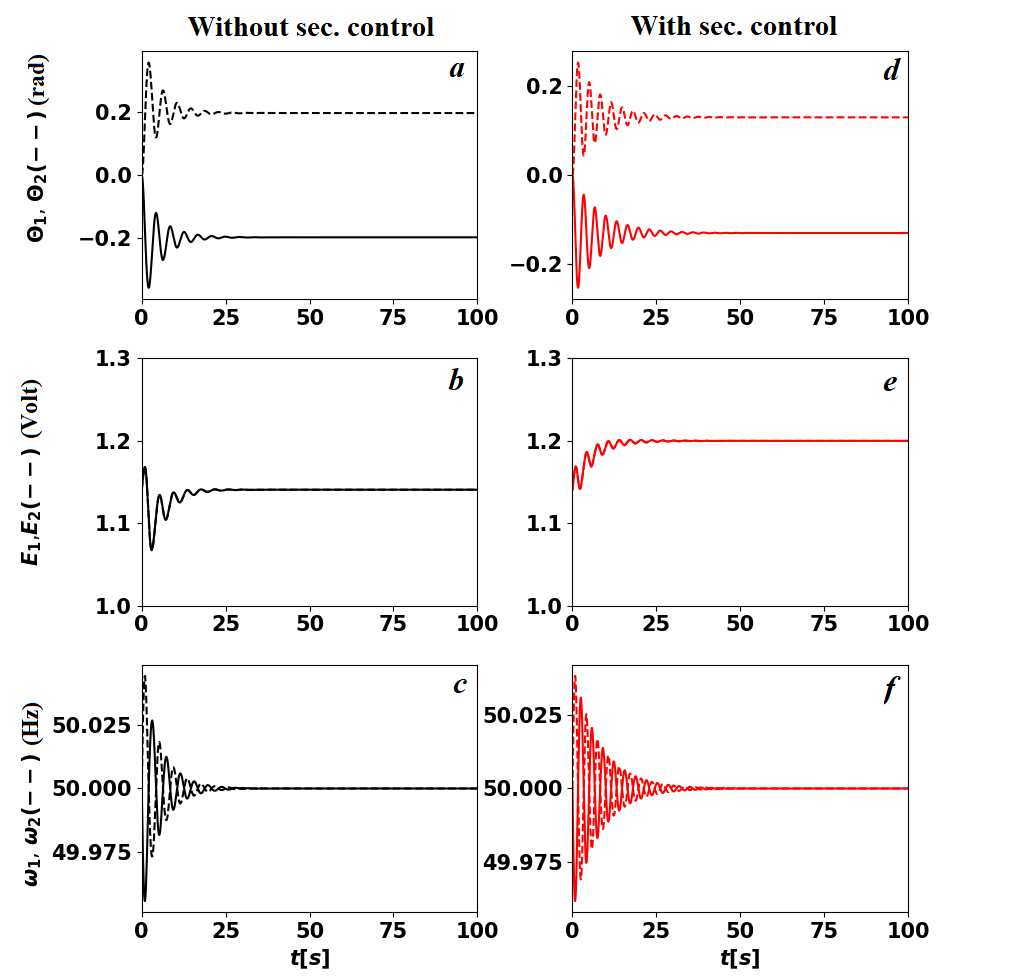}
\par\end{centering}
\caption{An unperturbed system converges to a steady state with and without control.
[Description of figure] parameters of simulation are the following: $\alpha_{1}=\alpha_{2}=0.2$, $T_{d,1}=T_{d,2}=2$,$E_{f,1}=E_{f,2}=1.$ , $E_{1}(0)=E_{2}(0)=1.14$, $\omega_{1}(0)=\omega_{2}(0)=50$ , $\theta_{1}(0)=\theta_{2}(0)=0$, $P_{dist}$=0, $\gamma_{1}=\gamma_{2}= 0$ (first column), $\gamma_{1}=\gamma_{2}= 1$ (second column)
}
\label{fig:withoutdist}
\end{figure}

The two node system considered here consists of a generator node of power $P_{1}>0$ and a consumer node of power $P_{2}<0$. The two nodes represent simply two interconnected power grids (or synchronous machines) with equal control parameters $\gamma_{1}=\gamma_{2}=\gamma$ and are described by the Eq. \eqref{full_3r_dmodel2} \revise{and parameters, such as $T_d$, inspired by literature values \cite{schmietendorf2014,auer2016}.} Initially, without disturbance, the two systems evolve till reaching a stable final state, in which the variables of each node tend to a steady state as shown in Fig.\ref{fig:withoutdist} without control (first column in black) and with control (second column in red).
One observes that in both cases the network is synchronized, meaning that the frequency at each node is equal to $\omega_{syn}= 50$ $Hz$ and the difference of phases between the connected node tends to a constant value. The final states reached by the voltages in the controlled case are sightly greater than the the ones of the uncontrolled case. This slight increase results from the effects of the secondary control parameter at each node, which affects the dynamics of the voltage through the angles. In fact, at a constant voltage, the secondary control is basically the governor which activates the online or offline substations, thereby increasing or reducing the power generated in order to balance the power in the system ($\omega=0$) \cite{Tchawou1, tchuisseu2018curing, tchuisseu2019reduction}. Thus, once the frequency is brought back to its nominal value, the phases in each node tend to constant values, which basically implies a constant difference of phases between connected nodes, hence constant voltage. Without control, the two nodes of the system evolve until sharing a constant amount of power and reach a power balanced state.
\begin{figure}[ht]
\begin{centering}
\includegraphics[width=1\linewidth]{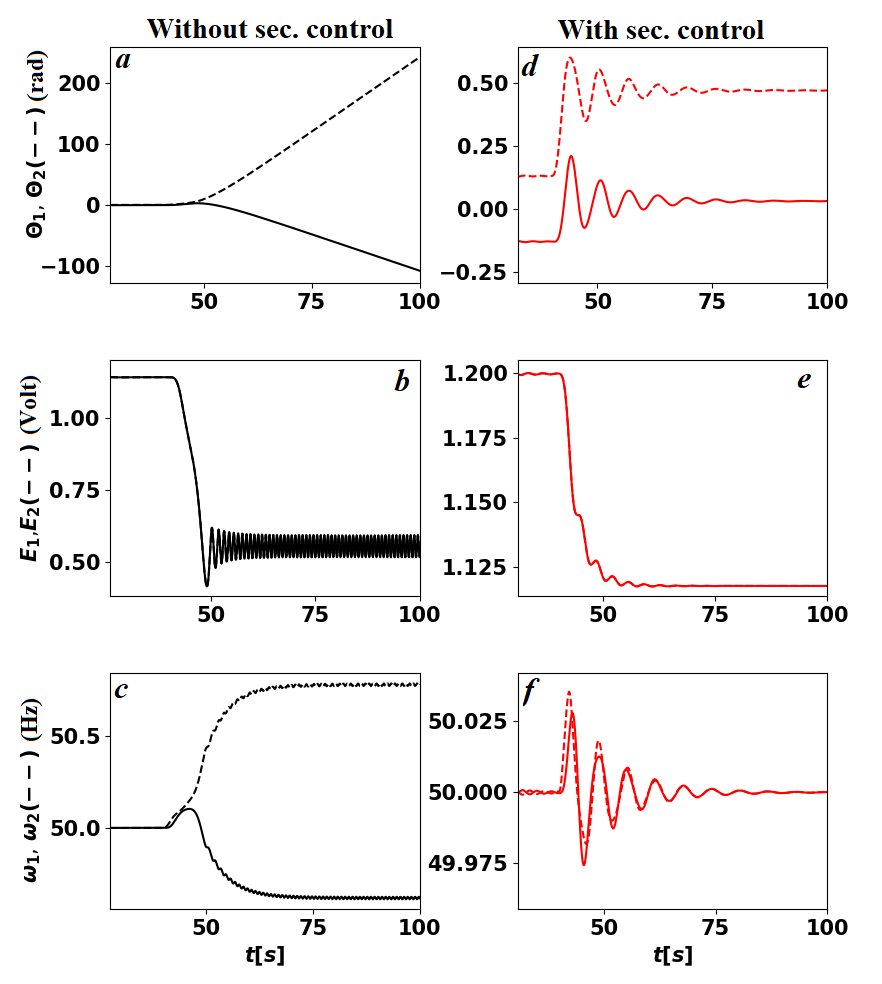}
\par\end{centering}
\caption{Secondary control stabilizes a 2 node system.
[Description of figure] parameters of simulation are the following: $\alpha_{1}=\alpha_{2}=0.2$, $T_{d,1}=T_{d,2}=2$,$E_{f,1}=E_{f,2}=1.$, $E_{1}(0)=E_{2}(0)=1.14$, $\omega_{1}(0)=\omega_{2}(0)=50$, $\theta_{1}(0)=\theta_{2}(0)=0$, $P_{dist}$=1, $\gamma_{1}=\gamma_{2}= 0$ (first column), $\gamma_{1}=\gamma_{2}= 1$ (second column)}
\label{fig:withdist}
\end{figure}
Next, we consider a perturbation to the power system: We assume that from time $t=40$ $s$ to $t=42$ $s$ , the power $P_{1}$ at node 1 experiences a gradual increase of power from 0 to $P_{dist}=1$, resulting in a sudden increase of its corresponding frequency, as shown in Fig. \ref{fig:withdist} (dashed lines) in the controlled (red) and uncontrolled (black) case. This perturbation leads to an instability of the uncontrolled case, visible by the emergence of periodic oscillation in the frequency and the voltages around states far away from their original states,while the corresponding phases are rising in opposite directions. 
On the other hand, the variables in the controlled system  (plot in red in Fig.\ref{fig:withdist}) converge to new steady states for the voltages and phases, while the corresponding frequencies return to their synchronous states as before the perturbation. Thus, the secondary control not only stabilized the frequencies but also stabilized the voltage, thereby avoiding  the system to drop into an unstable regime. Thus, the secondary frequency control acts as a damping for the voltages. This is a very similar behavior as the primary frequency control has on the frequency through the damping coefficient $\alpha$ on the frequency of the uncontrolled ($\gamma=0$) network. Hence, we can consider secondary frequency control acts somehow as a primary control for the voltage.

\begin{figure*}[ht!]
    \centering
    \includegraphics[width=0.9\linewidth]{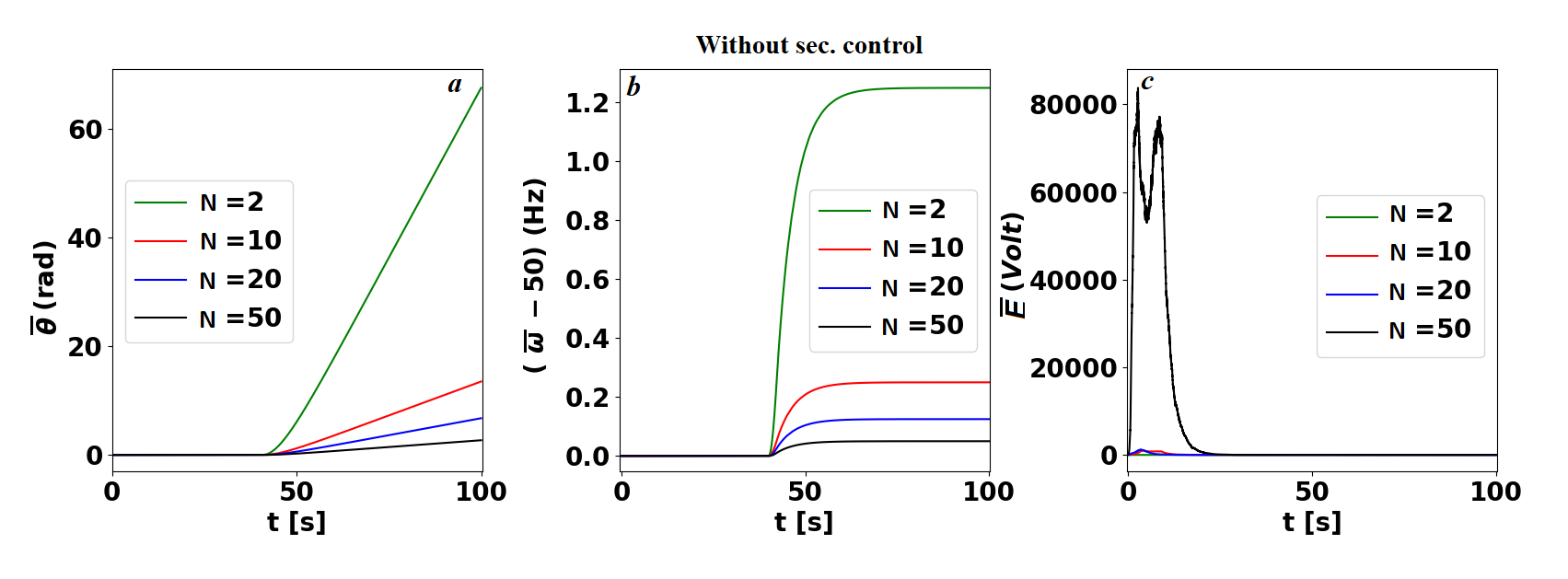}
    \includegraphics[width=0.9\linewidth]{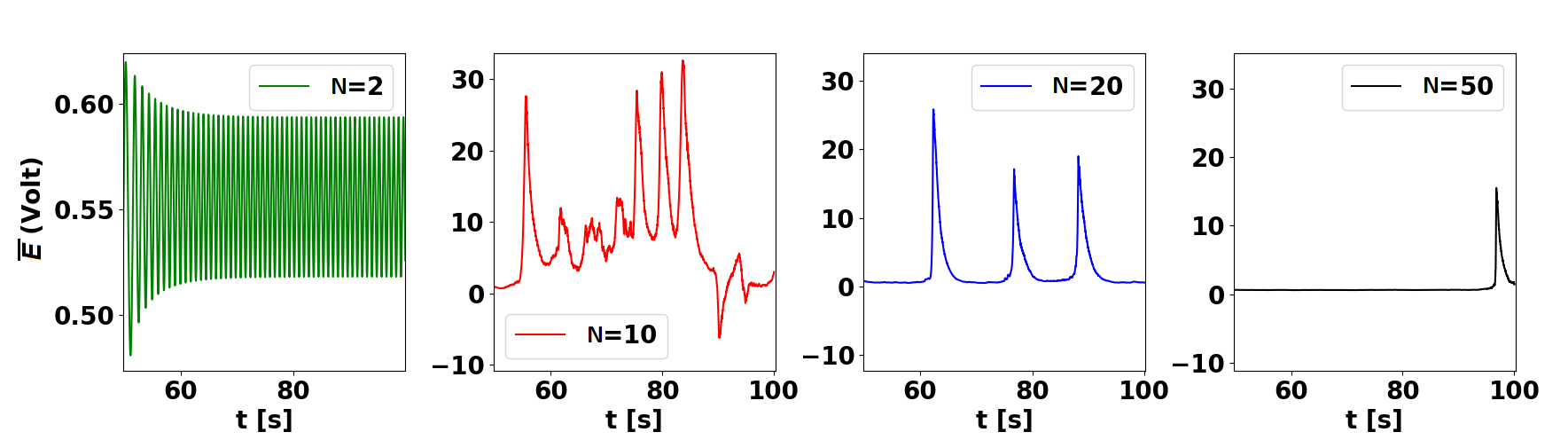}
    \caption{Uncontrolled networks display frequency instability and large voltage oscillations. Illustration of average of the phase (a), frequency deviation (b) and the voltage (c) as a function of the time for different size of the network, with the corresponding zoom of the voltage dynamics given at the bottom of the figure. Parameter values: $\gamma=0$, $\alpha=0.2$, $Pdist=1$, $X=1$, $E_{f}=1$, $E_{i}(t=0)=1.14$, $T_{d}=1$, $B_{0}=-0.8$ and $B_{1}=1$.
    \label{voltagedynwithoutcontrol}}
\end{figure*}

\begin{figure*}[ht!]
    \centering
    \includegraphics[width=0.9\linewidth]{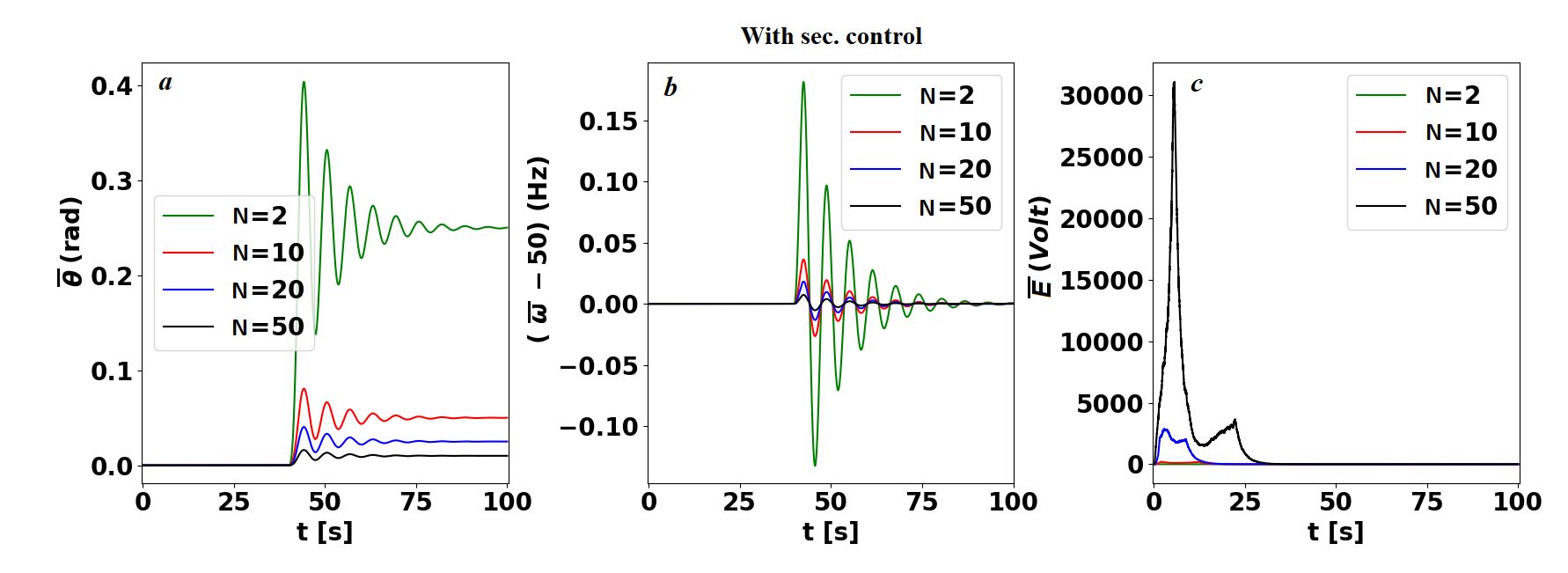}
    \includegraphics[width=0.9\linewidth]{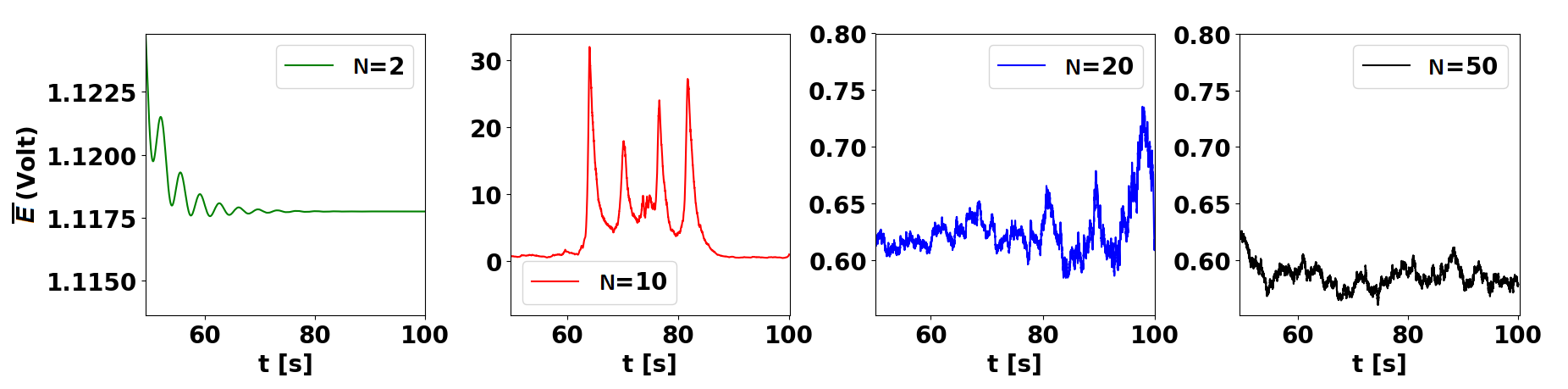}
    \caption{Secondary control stabilizes the frequency and voltage following initial oscillations. 
    Illustration of average of the phase (a), frequency deviation (b) and the voltage (c) as a function of the time for different size of the network, with the corresponding zoom of the voltage dynamics given at the bottom of the figure.
    Parameter values: $\gamma=1$, $\alpha=0.2$, $Pdist=1$}
    \label{voltagewithcontrolzoom}
\end{figure*}

\subsection{Bulk dynamics}
Moving towards the $N>2$, case, we now present the numerical results complementing the analytical bulk analysis. Thus, the question to be answered is the following: how does the secondary control affect the mean voltage dynamics in networks?
Thus, we consider the electric network, whose dynamics is governed by the Eq.~\eqref{full_3r_dmodel2} and we analyse the dynamics of the mean voltage and frequency for different network sizes. We consider for these purposes that we have an interconnected network constituted of $n$ nodes, which can represent $n$ interconnected isolated power grids or synchronous machines. 
\revise{So far, we assumed that the nodes are all-to-all-coupled, each node having equal power $P_{i}$ in absolute value, and the network without perturbation is power balanced  ($\sum_i {P_{i}=0}$). Such all-to-all-coupled networks naturally emerge after Kron reduction of any network topology \cite{Dorfler2013}.}
In all the studied scenarios, the system is perturbed  from the time $t=40$ $s$ to $t=42$ $s$ by gradually increasing the power at a single node from $0$ to $P_{dist}$.\\

First, we consider the uncontrolled ($\gamma=0$) case and plot the deviation of the mean phase, frequency and voltage as function  of time for different network size, and without secondary control in Fig.~\ref{voltagedynwithoutcontrol}. 
We observe a clear agreement with the analytical predictions given by the expression of the frequency deviation in Eq.~\eqref{dev}, meaning that the mean frequency for a long time tends to a constant. Since we assumed that the primary control parameter/the inertia of each power plant is constant, we observe that the mean of the frequency deviation decreases when the size of the network increases. The corresponding curve of the average voltage is plotted at the third column of the first row. Complementing the analytical results, we observe large transient voltage dynamics, which increase in amplitude with network size. 
To better observe the dynamics of the mean value of the voltage towards the end of our simulation window, we provide zooms in the lower row of Fig.~\ref{voltagedynwithoutcontrol} for each size $n$ of the network. 
For the two nodes system, the average frequency tends to a constant value. The mean voltage conversely is oscillating around a stable value $\overline{E} \approx 0.55$ $Volt$. For larger values of $n$ on the other hand, the mean voltage is fluctuating and displays large periodic peaks. These regular peaks persist throughout the running time of our simulation and are observed for $N$ $\textbf{=}$ $10$ and $N$ $\textbf{=}$ $20$. For $N$ $\textbf{=}$ $50$, after the transient, the mean voltage mostly fluctuates around a constant value and only displays one peak, likely pointing to a longer periodicity of these peaks due to the larger network. The small fluctuations observed are probably due to the difference of phases existing between the connected nodes. In this uncontrolled studied case, the system is clearly not a synchronous stable electric network.

Now, we include secondary control ($\gamma>0$) and repeat the same simulations as before. Fig.~\ref{voltagewithcontrolzoom} (top) shows the time evolution of the mean frequency and the mean voltage of the considered network in presence of the secondary control. First, we notice that the mean frequency deviation tend to zero for every network size, as we predicted in our linear stability analysis. Hence, we focus on the mean voltage dynamics, for which we could not derive an equality from the Eq.\eqref{eq3} but only constrained its range of fluctuations.  Thus, only these numerical simulation can tell us how the voltage evolves with time. Similar to the frequency,  after an initial transient phase, the mean voltage tends to a constant value for all the considered sizes of the network. 

Again, to highlight the voltage dynamics towards the end of our simulation window, we provide a zoom of the voltages in the second row of  figure~\ref{voltagewithcontrolzoom}. For the two node system, we observe that the mean voltage tends to steady state, and it is not oscillating as in the corresponding uncontrolled case. Thus, the secondary control clearly stabilizes the voltage. In fact, without secondary control, the phase of each node is increasing continuously with the time (see Fig.~\ref{fig:withdist}), leading to the oscillation of the voltage. The secondary frequency control on other hand stabilizes the phases, hence the cosine function in the expression of each voltage becomes a constant, leading to constant voltage at each node. For larger networks, it may happen that due to the power flow between connected nodes, the phases slightly vary, leading at some times to the fluctuations of the voltage. This justifies some rare peaks observed in the mean voltage. Nevertheless, the magnitude of the voltage and its peaks in the controlled network mostly remains lower than the one in the uncontrolled network. In addition, we observe that the steady state reached in the end decreases when the network size increases.
\subsection{Relaxation time}

\begin{figure}[ht!]
    \centering
    \includegraphics[width=0.9\linewidth]{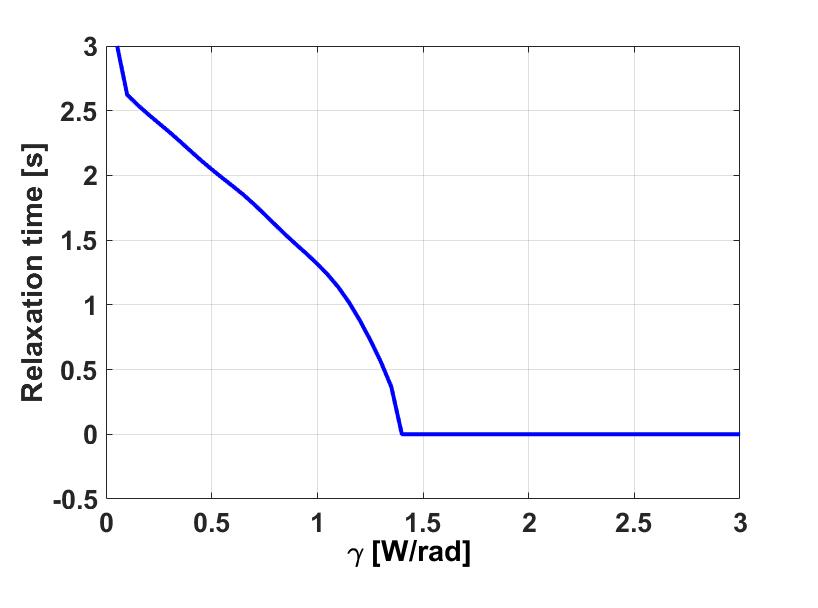}
    \caption{Secondary control stabilizes the system by reducing the return time. We plot the return time of the mean voltage when the control parameter increases in the two node system. The parameters of simulation are the following: $\alpha_{1}=\alpha_{2}=0.2$, $T_{d,1}=T_{d,2}=2$, $E_{f,1}=E_{f,2}=1$, $P_{dist}=1$ $N=2$ nodes. 
    }
    \label{fig:return}
\end{figure}

We have shown in the previous analysis how the dynamics of the phase, the frequency, the voltage as well as their corresponding mean values evolve with the secondary parameter $\gamma$, once the network is perturbed. In fact, once perturbed from its stable state, the variables of the system vary until they reach new steady states, often different from their original steady states without perturbation. Now, we investigate how long it takes the system to relax to its (new) steady state, by computing the \textit{return time or relaxation time}. According to the bulk dynamics of the network, the return time of the frequency and phase average depends strongly on the damping of the system, and slightly on the secondary control parameter, which impacts mostly the oscillatory regime of these variables. If the return time of the mean frequency and phase can be obtained explicitly, this cannot so easily be achieved for the mean value of the voltage. Hence, we again use numerical computations to obtain the return time for different control parameters. \\
In particular, we consider the previously described 2 nodes system, including voltage dynamics. For this system, we aim to determine the evolution of the return time as a function of the secondary control parameter $\gamma$. The network is perturbed as previously, by gradually decreasing the power at the node $1$ from $P_1$ to $P_{1}+P_{dist}$ ($P_{dist}=-1$) from time t=$40 s$ to t=$42 s$ and all other parameters also remain unchanged. The relaxation time or return time is numerically computed by recording the time taken by the voltage at a node to regain a steady state after perturbation. Thus, we define by $\overline{E}(t)$ the mean voltage at the time $t$ and $\overline{E}(t-T)$ the mean at the time $t=t-T$, where $T$ is a characteristic time. We define also by $\xi$ the numerical tolerance. The mean voltage is considered stable after perturbation of the network if and only if $|\overline{E}(t)-\overline{E}(t-T)|\leq$ $\xi$. Note that we focus here on the relaxation and therefore of the voltage, not the frequency, as we are interested in quantifying the impact of frequency control on voltage stability.
The so computed return times decrease with increasing secondary control parameter $\gamma$, see Fig.~\ref{fig:return}. First, we note that for any non-zero value of the secondary control parameter $\gamma$, the system after perturbation always regains a steady state, i.e. the secondary frequency control guarantees stability and return. Secondly, we observe that the return time is reduced by increasing the control amplitude, approaching zero for sufficiently large control $\gamma$. This means that the secondary frequency control is strong enough to immediately compensate the introduced disturbance and the system never leaves its original fixed point.

\begin{figure*}[ht]
\begin{centering}
\includegraphics[width=1.0\linewidth]{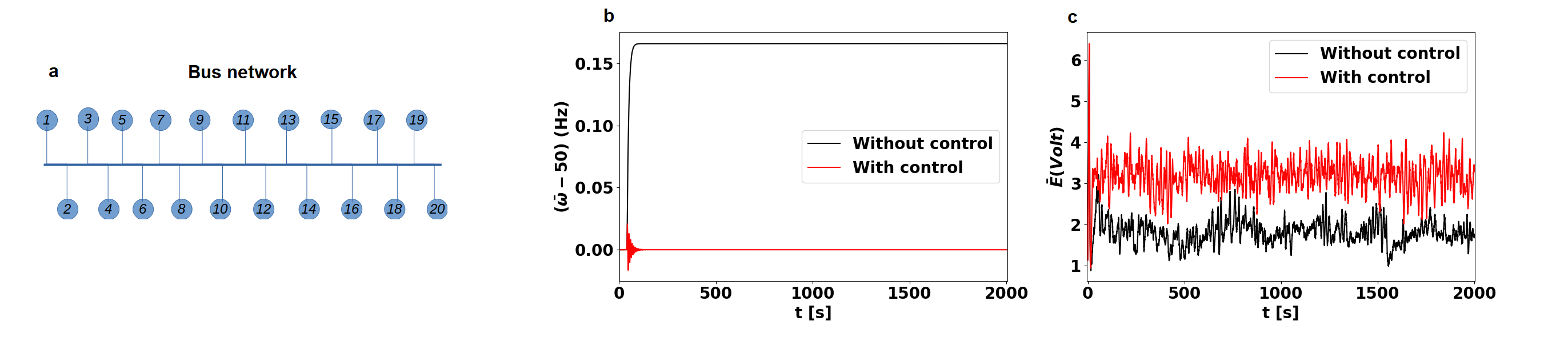}
\par\end{centering}
\caption{\revise{Exploring heterogeneous parameters and non-all-to-all-coupling.
Illustration of the topology (a), the average of the frequency deviation (b) and the voltage (c) as a function of the time for the controlled (black) and controlled (red) network.
Parameter values: $P=[-0.4, 0.53, -0.51, 0.56, 0.52, 0.48,  -0.55, -0.45, 0.491, 0.509, -0.482, -0.518, -0.46,-0.64, 0.42, 0.58, -0.5, 0.5, 0.35, -0.45]$, primary control $\alpha_i=0.2|P_i|$, secondary control $\gamma_i=|P_i|$, while $P_{dist}$, $X$ and $B$ are chosen as in the other simulations.
}}\label{fig:heterogeneous}
\end{figure*}

\revise{
\subsection{Case study: Heterogeneous parameters and network influence}
So far we have assumed networks that are all-to-all-coupled and have homogeneous parameters, i.e. identical loads, generation and control values. We used this simplification to derive analytical results. To demonstrate that our results are in principle also applicable to more general systems we consider one non-all-to-all-coupled system with heterogeneous parameters, see Fig.~\ref{fig:heterogeneous}a. In particular, we simulate the dynamical behavior of 20 nodes connected to a common bus with power values $P$ randomly drawn between -0.7 and 0.7. We assume nodes with large absolute value of $P$ to provide more control power and hence set the control values proportional to the absolute power value: $\alpha_i=0.2|P_i|$, secondary control $\gamma_i=|P_i|$, where $|...|$ denotes the absolute value. 
Analogue to the earlier analysis, we observe that the average frequency can only be restored once secondary control is used, see Fig.~\ref{fig:heterogeneous}b. Meanwhile, the voltage dynamics is not controlled and oscillates (Fig.~\ref{fig:heterogeneous}c) since we have more than two nodes, consistent with our analytical results.  How exactly non-homogeneous parameters and network topology affect the dynamics and controlability of both frequency and voltage is beyond the scope of this study.
}

\section{Conclusion}
This article has investigated the effects of the secondary frequency control on the voltage dynamics of electric networks of different sizes. We considered an \revise{simple networks, both all-to-all-coupled as well as a bus topology.}
Our analytical linear stability analysis of the network has shown that the secondary control can guarantee the stability of the network. In addition, considering the network as a simplified bulk, we have demonstrated that the stability of the mean phase and frequency are independent of the mean voltage of the network. On the other hand, the mean voltage does depend on the nodes' phases. The different numerical simulations computed for the perturbed network and in presence of the secondary control have shown that, the secondary control actually plays the role of a primary control for the voltage. The frequency secondary control after perturbation stops the variation of the voltage and stabilizes it to a new steady value different to the original one. 

Our results showcase how voltage stability and secondary frequency control should be considered in other power system stability analyses: Including only primary frequency control might stabilize the frequency but does not guarantee any voltage stability. Hence, the voltage should then explicitly be considered when assessing stability in power systems. Meanwhile, if secondary frequency control is applied, the voltage stability is no longer an immediate concern and might be neglected, especially if only the short-term stability is of interest.

In the future, it would be interesting to complement the secondary frequency control, which acts as an effective "primary voltage control" by a "secondary voltage control" to bring the voltage back within its operational boundaries.  \revise{Furthermore, how precisely network topology and heterogeneous parameters affect the voltage and frequency stability and return times still remains a mostly open question for now. Similarly, the choice of specific parameters, such as $T_d$ and $X_i$ should be investigated further.} Finally, our linear stability and return time analysis could be supplemented by a detailed analysis of the basin of attraction \cite{Hellmann2016}.

\subsection*{Acknowledgments}
This project has received funding from the European Union’s Horizon 2020 research and innovation programme under the Marie Sklodowska-Curie grant agreement No 840825.\\
E.B.T.T. and P.P. acknowledge funding from the European Regional Development Fund under Grant No. CZ.02.1.02/0.0/0.0/15003/0000493, "CENDYNMAT - Centre of Excellence for Nonlinear Dynamics Behaviour of Advanced Materials in Engineering" and the Academy of Sciences CR under Grant Strategy AV21, VP03: "Efficient energy conversion and storage, Vibrodiagnostics of rotating blades of rotary machines in power engineering.\\
E.B.T.T., D.G., and P.C. acknowledge funding from the Ministerio de Ciencia e Innovaci\'on (Spain), the Agencia Estatal de Investigaci\'on (AEI, Spain), and the Fondo Europeo de DesarrolloRegional (FEDER, EU) under Grant No. PACSS (RTI2018-093732-B-C22) and the Maria de Maeztu program for Units of Excellence in R\&D (No. MDM-2017-0711). E.B.T.T. also acknowledges the fellowship from the AEI and MINEICO, Spain under the FPI program(No. FIS2015-63628-CZ-Z-R).

\subsection*{Competing interests}
The authors declare no competing interests.

\bibliographystyle{apsrev}
\bibliography{references}

\appendix
\label{appendixx}
\section{Eigenvalues of sum of matrices}
To compute the eigenvalues of the sum of two matrices, as used in this paper, we follow this simple line of reasoning: The eigenvalue equation of matrix $A$ are obtained by:
\begin{equation}
    \det{\left(A-\lambda I\right)}=0,
\end{equation}
with identity matrix $I$ and where $\lambda_1, \lambda_2, ..., \lambda_N$ are the eigenvalues of $A$. Now suppose we add a diagonal matrix to $A$, i.e. $B=A+\gamma I$, then the eigenvalue equation reads
\begin{equation}
\begin{aligned}
    \det{\left(B-\tilde{\lambda} I\right)}&=& 0, \\
    \det{\left(A+\gamma I -\tilde{\lambda} I\right)}&=& 0,\\
    \det{\left(A- \left(\tilde{\lambda} - \gamma \right) I\right)}&=& 0,\\
    \det{\left(A- \lambda I\right)}&=& 0,
\end{aligned}
\end{equation}
i.e. the eigenvalues $\tilde{\lambda}$ of $B=A+\gamma I$ are given as $\tilde{\lambda}_k=\lambda_k+\gamma$.
So if the eigenvalues are ordered as $\lambda_1 \leq \lambda_2 \leq ...$ and $\lambda_1<0$ is the smallest eigenvalue, we can choose any $\gamma>\lambda_1$ so that $B=A+\gamma I$ has only positive eigenvalues, i.e. it the matrix $B$ is positive definite.

The interested reader might consult more general results on the eigenvalues of two Hermitian matrices \cite{horn1962eigenvalues}.

\end{document}